\documentclass{article}
\usepackage[margin=1.2in]{geometry}  
\usepackage[export]{adjustbox} 
\usepackage{amssymb}

\graphicspath{ {./materials/figures/} }

\usepackage[utf8]{inputenc}  
\usepackage[T1]{fontenc}
\usepackage{subcaption} 
\usepackage{amsmath} 
\newcommand{\D}{\mathrm{d}} 
\newcommand{\E}{\mathbb{E}} 

\usepackage[compress,sort,numbers]{natbib}

\usepackage{xcolor} 
\usepackage{hyperref}
\hypersetup{
    colorlinks=true,
    linkcolor=blue,
    citecolor=gray,
    filecolor=magenta,      
    urlcolor=cyan,
}

\usepackage[noblocks]{authblk} 

\author[1,2]{Arthur Prat-Carrabin} 
\author[3]{Florent Meyniel}
\author[4,5]{Misha Tsodyks}
\author[2,4,6,7,$\star$]{Rava Azeredo da Silveira}

\affil[1]{Department of Economics, Columbia University, New York, USA}
\affil[2]{Laboratoire de Physique de l’École Normale Supérieure, ENS, Université PSL, CNRS, Sorbonne Université, Université de Paris, France}
\affil[3]{Cognitive Neuroimaging Unit, Institut National de la Santé et de la Recherche Médicale, Commissariat à l’Energie Atomique et aux énergies alternatives, Université Paris-Saclay, NeuroSpin center, 91191 Gif/Yvette, France}
\affil[4]{Department of Neurobiology, Weizmann Institute of Science, Rehovot, Israel}
\affil[5]{The Simons Center for Systems Biology, Institute for Advanced Study, Princeton, USA}
\affil[6]{Institute of Molecular and Clinical Ophthalmology Basel}
\affil[7]{Faculty of Science, University of Basel}
\affil[$\star$]{\textit{email: rava@ens.fr}}

\title{Biases and Variability from Costly Bayesian Inference}
\date{}							

\begin{document}
\maketitle

\begin{abstract}
When humans infer underlying probabilities from stochastic observations, they exhibit biases and variability that cannot be explained on the basis of sound, Bayesian manipulations of probability. This is especially salient when beliefs are updated as a function of sequential observations. We introduce a theoretical framework in which biases and variability emerge from a trade-off between Bayesian inference and a cognitive cost of carrying out probabilistic computations. We consider two forms of the cost: a precision cost and an unpredictability cost; these penalize beliefs that are less entropic and less deterministic, respectively. We apply our framework to the case of a Bernoulli variable: the bias of a coin is inferred from a sequence of coin flips. Theoretical predictions are qualitatively different depending on the form of the cost. A precision cost induces overestimation of small probabilities on average and a limited memory of past observations, and, consequently, a fluctuating bias. An unpredictability cost induces underestimation of small probabilities and a fixed bias that remains appreciable even for nearly unbiased observations. The case of a fair (equiprobable) coin, however, is singular, with non-trivial and slow fluctuations of the inferred bias. The proposed framework of costly Bayesian inference illustrates the richness of a `resource-rational' (or `bounded-rational') picture of seemingly irrational human cognition.
\end{abstract}

\section{Introduction}

While the faculty of rational
thinking defines, at least to an extent, our human nature, it suffers from a
remarkably long list of so-called `cognitive biases'---systematic deviations
from rational information processing and behavior \cite{wikipedia_biases}.
Much as optical illusions are informative of the neurobiological mechanisms of
vision \cite{hubel1995eye}, one expects cognitive biases to reveal aspects of
the algorithmic basis of cognition and behavior \cite{baron2000thinking}. A
notable category of biases comprises those that govern the way we manipulate
probabilistic quantities: these biases affect our inference of the probability
of events, our decision-making process, and more generally our behavior in
situations where stimuli obey (seeming or unknown) stochastic rules
\cite{wendt2012utility,gilovich2002heuristics,hilbert2012toward,group2014evolution,summerfield2015humans,Meyniel2016,Gonzalez1999,Zhang2020,Hertwig2004}.

In such situations, human subjects violate a normative predicament viewed
by the experimenter as the rational one (e.g.,\textquotedblleft maximize the
number of correct responses per unit time, in a given task\textquotedblright)
\cite{ma2012organizing}. In a task involving random stimuli, a
subject's brain may fail to manipulate the probabilities it infers from the
stimuli in a sound (Bayesian) and precise mathematical manner, and, as a result,
predictions or decisions based upon the inferred probabilities may suffer from
biases. Several origins of this phenomenon have been proposed. One common
suggestion is that the brain manipulates probabilities in the correct,
Bayesian way, but that the outcome of the inference process is altered by the
choice of an `erroneous' prior probability
\cite{weiss2002motion,stocker2006noise,ma2012organizing}.
In other words, the subject may reason correctly, but base the reasoning upon incorrect prior beliefs.
It is also possible
that the brain carries out Bayesian calculations, but with noisy processing units (neurons)
that yield stochastic, approximate responses \cite{Khaw2017} or posteriors \cite{acerbi2014origins,drugowitsch2016computational,Prat-Carrabin2021}.
Alternatively, the brain may simply not be concerned with a mathematically sound
manipulation of probabilities, but rather may operate with the use of a set of
heuristic algorithms \cite{gilovich2002heuristics,gigerenzer2011heuristic}.

Here, we propose another, but possibly complementary, cognitive mechanism for the
emergence of biases in the inference of probabilities.
We assume that in order to perform Bayesian inference, the brain needs to represent probability distributions, but bears a computational cost to do so.
When inferring a probability, $p$, based on a sequence of observations, the Bayesian observer updates a probability distribution, $P(p)$, after each new observation.
For the brain, however, producing a representation of a probability distribution is presumably an operation that comes with a metabolic cost, or that is subject to cognitive constraints. The brain, thus, might prefer to choose an alternative distribution, $\hat P(p)$, that is less costly to represent, but that still captures much of the behaviorally relevant structure in the Bayesian posterior, $P(p)$. We formalize this compromise as an optimization problem, in which the objective of choosing a posterior that is close to the optimum competes with the cost of the posterior. We examine two cost functions, that capture two natural assumptions about the cognitive constraints at play. The \textit{precision cost}, first, follows from the assumption that a more precise encoding by the brain requires greater metabolic resources: in general, in information-theoretic terms, a larger number of bits is needed to resolve finer uncertainty. Mechanistically, a more precise encoding of information in a brain area may require the use of more neurons or more spikes, involving a greater metabolic cost.
The \textit{unpredictability cost}, second, penalizes beliefs that imply more uncertain (or less deterministic) environments, which are more difficult for the brain to cope with and represent an ecological challenge.

We examine the ways in which these two costs impact inference.
For the sake of simplicity, we focus on the scenario of an observer facing a series of binary signals (i.e., $p$ is the parameter of a Bernoulli distribution). In contrast to other studies in which the statistics of the inferred variable undergo changes \cite{Gallistel2014,Khaw2017,Prat-Carrabin2021}, here they are constant.
Yet because of the presence of a cost, the subject's estimate of $p$ will generally be biased, and the posterior, in some cases, does not even converge. In paying special attention to an equiprobable environment ($p=1/2$) --- the case of a fair coin ---, we show the sometimes surprising dynamics of the inference in this singular case. Our study proposes a fresh theoretical understanding of the biases that the humans exhibit when carrying out inference about probabilities.

\section{Results}

\subsection{Costly Bayesian inference}
We propose to regularize Bayesian inference by a cost on probability distributions.
We reformulate the inference procedure as a trade-off
between Bayesian updating of the probability distribution and a cost term
that penalizes some distributions more than others (we consider two natural examples of possible costs, below).
While this modification of the
Bayesian procedure may appear harmless, and does not carry with it any
explicit bias, we show below that it may alter profoundly the outcome of the
inference of probabilities. We emphasize that our aim, here, is not to claim
that humans carry out the altered Bayesian inference that matches precisely
our mathematical formulation below, nor that our prescription fits behavioral
data better than another model; rather, we would like to present the idea of
regularized Bayesian inference in which the trade-off emerges from a cognitive constraint
as a possible, additional ingredient among a number of rationalizations
of cognitive biases.

In this spirit, we consider the simplest possible scenario of online
inference: successive flips of a (fair or biased) coin, from which a model
subject is asked to infer the probability of observing `head', $p$, (or
`tail', $1-p$) in the upcoming flip. 
(Equivalently, the model subject infers the `bias' of the coin, i.e., the extent to which the probability differs from 0.5.)
We assume that the model subject carries out online inference by updating a
probability distribution, $\hat P_{N}\left(  p\right)$, at successive coin flips
indexed by $N$.
A natural choice of distribution, after each coin flip, is the posterior prescribed by Bayes' rule, $P_N$; and our model subject would select this distribution if it were not for the cognitive cost, $C[P]$, associated with any distribution $P$. Instead, the model subject chooses a compromise between bearing the cost, $C[\hat P_N]$, of the inferred posterior, $\hat P_N$, and minimizing the discrepancy between the inferred posterior and the Bayesian posterior, $P_N$, as measured by a function that we denote by $D(\hat P_N, P_N)$. Specifically, the model subject minimizes the loss function

\begin{equation}\label{eq:loss}
L = D(\hat P_N, P_N) + \lambda C[\hat P_N],
\end{equation}
where $\lambda \ge 0$, the only parameter in the theory, determines the strength of the regularization from a cognitive cost. We define the distance as the Kullback-Leibler divergence between the inferred distribution and the Bayesian posterior, $D_{KL}(\hat P_N || P_N)$. A similar loss function has been proposed in the literature, using the Kullback-Leibler divergence as a distance metric, in modeling the sub-optimality of cognition \cite{Icard2015}. 
We will show that even a cost that does not carry with it any explicit bias may alter the outcome of the inference qualitatively. Depending upon the nature of the cognitive toll borne by the brain to represent a probability distribution, and depending on the coin's degree of bias (or lack thereof), the inference process of the model subject may converge to an under- or an over-estimation of the bias, or may not converge at all. We obtain a diversity of behavioral patterns --- some, counter-intuitive --- by examining just two costs that derive from natural hypotheses on the constraints at work when the brain manipulates probability distributions. We now present these costs and the behaviors they yield.

\subsection{Inference of probability under a precision cost}
A natural hypothesis is that costly distributions are the ones that reduce the uncertainty on the inferred latent parameter (here, $p$). Specifically, for a distribution, $P(p)$, we propose a `precision cost' equal to the negative of its Shannon entropy, i.e.,

\begin{equation}\label{eq:icost}
C[P] \equiv - H[P] = \int P(p) \ln P(p) \D p.
\end{equation}
Under this hypothesis, the uniform distribution is the least costly distribution (with cost zero), and we note that this cost function is equal to the Kullback-Leibler divergence between the uniform distribution and the inferred distribution, $P$. More `precise' distributions, that are concentrated around some value, come with a higher cost.
The distribution that minimizes the loss function (Eq. (\ref{eq:loss})), with this cost, is proportional to the Bayesian posterior raised to an exponent:

\begin{equation}
\hat P_N(p) \propto P_N(p)^{\frac{1}{\lambda+1}}.
\end{equation}
In the absence of cost ($\lambda=0$), the inferred distribution coincides with the Bayesian posterior;
with an infinitely large cost ($\lambda \to \infty)$, the solution is the uniform distribution;
and finite, positive values of $\lambda$ yield intermediate distributions. After a new coin flip is observed, the updated Bayesian posterior is proportional to the product of the prior and the likelihood, and thus the updated inferred posterior is proportional to these two quantities, raised to the exponent $1/(\lambda+1)$. A similar form of the posterior has been posited to capture human biases in probability estimations \cite{Benjamin2019}; here, it is obtained as the solution of an optimization argument.

Before the first coin flip, we assume that our model subject holds a uniform prior, i.e., $\hat P_0(p)=1$. In this case, the probability distribution after the $N$th flip is a Beta distribution,

\begin{equation}\label{eq:beta_posterior}
\hat P_N(p) \propto p^{\hat n_1} (1-p)^{\hat n_0},
\end{equation}
where $\hat n_1$ and $\hat n_0$ are exponentially filtered counts of `heads' and `tails',
\begin{align}\label{eq:n0n1}
& \hat n_1 = \sum_{k=0}^{N-1} \Big(\frac{1}{\lambda+1}\Big)^{k+1} x_{N-k},  \\
\text{ and } & \hat n_0 = \sum_{k=0}^{N-1} \Big(\frac{1}{\lambda+1}\Big)^{k+1} (1-x_{N-k}),
\end{align}
where $x_i$ is 1 if `head' occurs at the $i$th flip, and 0 if it is `tail'.

\begin{figure}[htbp]
\includegraphics[max height=0.9\textheight, max width=\textwidth]{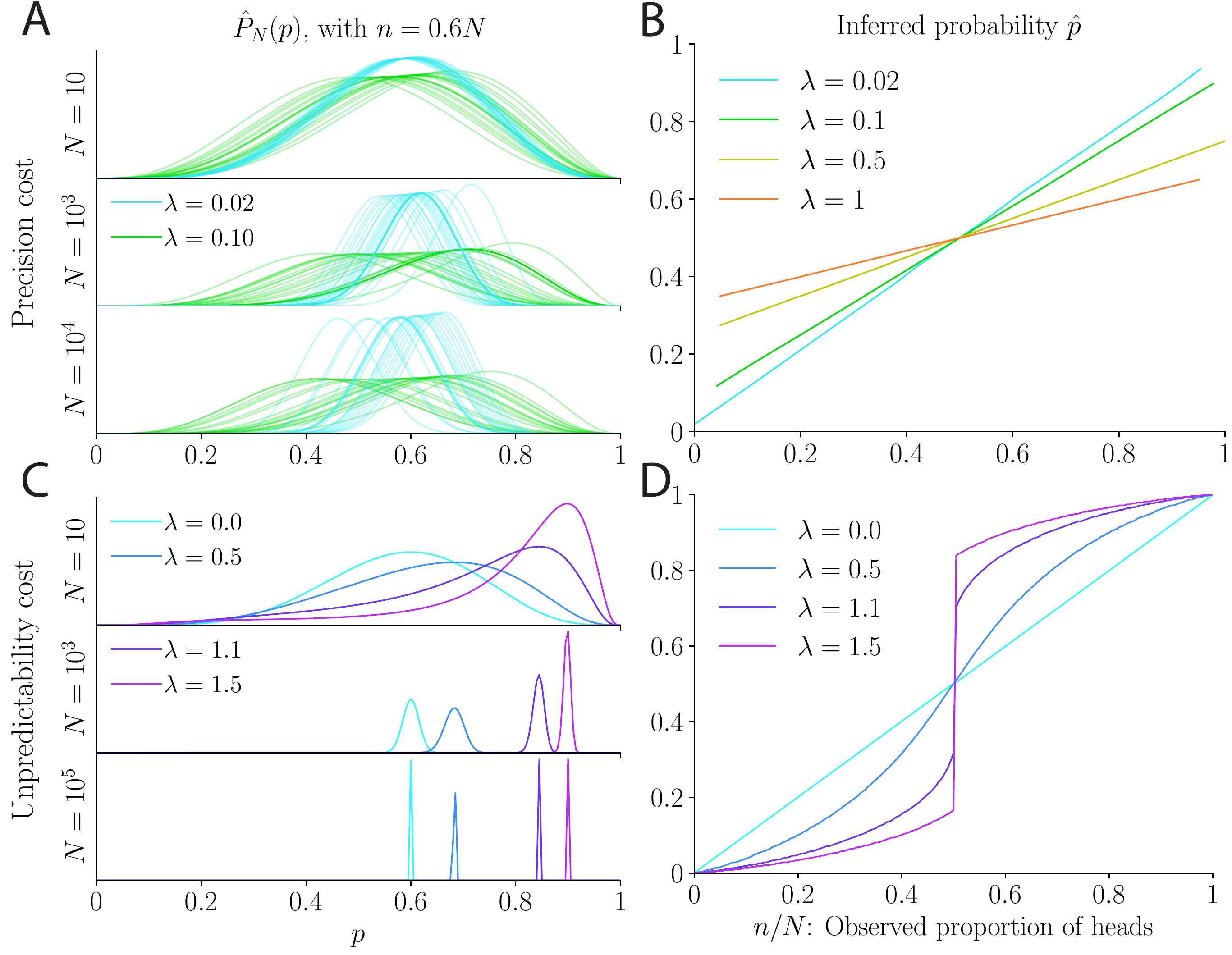}
\caption{\textbf{Biased coin: probability inference under a precision cost (A,B) and an unpredictability cost (C,D).}
\textbf{A.} Inferred distribution, $\hat P_N(p)$, for the precision cost with $\lambda=.02$ and .10, after $N=10, 1000$, and 10,000 coin flips, and 60\% of `heads', in 20 sequences of coin flips. Different sequences result in different posteriors, and a larger cost yields wider posteriors. As more evidence is accumulated, the width of the posterior decreases at first, but plateaus at large $N$.
\textbf{B.} Average estimate of the model with the precision cost, $\hat p$, as a function of the proportion $n/N$ of `heads' observed.
\textbf{C.} Inferred distribution $\hat P_N(p)$ for the unpredictability cost, after $N=10, 1000$, and 100,000 coin flips and 60\% of `heads'. With this cost, given the number of coin flips, $N$, the ratio $n/N$ fully determines the posterior. As more evidence is accumulated, the posterior is more peaked around $\hat p$.
\textbf{D.} Inferred estimate of the model with the unpredictability cost, $\hat p$, as a function of the proportion $n/N$ of `heads' observed. For $\lambda > 0$, the cost exacerbates the true bias of the coin. For $\lambda > 1$, at exactly $n=N/2$ there are two local maxima. When $n$ fluctuates around $N/2$, the inferred probability will switch from one maximum to the other (see Figs. \ref{fig-fair} and \ref{fig-randomwalk}).
}
\label{fig-biased}
\end{figure}

In these counts, the outcomes of past coin flips are gradually `forgotten'; as a result, the posterior does not converge as more evidence is accumulated, but it fluctuates, instead, as a function of the recent history of coin flips. Hence, although the posterior of a perfectly Bayesian observer would be entirely determined by the total number of flips, $N$, and the number of `heads' among them, $n$, for our model subject different orders of `heads' and `tails' in the sequence result in different posteriors (Fig. \ref{fig-biased}A).
In addition, at long times the variance of the posterior is controlled by the strength of the cost, $\lambda$ (and by the probability, $p$).
Even after a large number of coin flips, the width of the posterior does not vanish; it is an increasing function of $\lambda$  (Fig. \ref{fig-biased}A; see Methods for an approximate expression of the variance of the posterior as a function of $\lambda$). Furthermore, the mean of the posterior, $\hat p$,
varies as a function of the recent history of flips; on average, however, it is a biased estimate of the probability, except in the case of a fair coin (a case we examine more closely below). With a biased coin, large probabilities are underestimated, i.e., $p>\E [\hat p] >1/2$, and small probabilities are overestimated, i.e., $p< \E [\hat p] <1/2$ (the expectations are taken over the space of all sequences, or, equivalently, over a very long sequence). In other words, the model subject underestimates the bias of the coin, and the more so for a stronger cost (Fig. \ref{fig-biased}B). We emphasize that we did not start by positing the exponentially filtered counts of the observations (Eq. (\ref{eq:n0n1})), or by making a related assumption on the limited memory of the model subject. The decaying counts result naturally from the assumption of a precision cost, from which also follow, thus, the underestimation of the coin's bias, and the fluctuations of the posterior.

\subsection{Inference of probability under an unpredictability cost}
We considered, above, the hypothesis that more concentrated posteriors entail a higher computational toll than broader ones. We now turn to a cost that varies not with the entropy of the posterior, but with that of the inferred probability. More precisely, as decisions are more easily made when the (inferred) environment is more predictable, we examine the hypothesis that the brain favors the probabilities that have more predictive power, and penalizes those that imply, conversely, unpredictability in the environment. We quantify the degree of unpredictability
of a coin flip by the entropy of the Bernoulli random variable corresponding to the coin flips,

\begin{equation}
H(p)  =-p\ln(p) - (1-p)  \ln(1-p),
\end{equation}
and we define the cost of a probability distribution, $P(p)$, as the average of this entropy over the posterior distribution, i.e.,

\begin{equation}\label{eq:ucost}
C[P] = \mathbb{E}_P [ H(p) ] = \int H(p) P(p) \D p.
\end{equation}
(While the precision cost is a function of the entropy of the \textit{posterior}, $H[P]$, the unpredictability cost is a function of the entropy implied by the \textit{probability}, $H(p)$, averaged over the posterior.)

Online Bayesian inference, when regularized by this cost (i.e., when the quantity in Eq. (\ref{eq:loss}), using this cost, is minimized), yields a probability distribution of the form ${\hat P_{N}(p)  \propto[
\varphi_{n/N}(p)]  ^{N}}$ (up to a normalization factor) at
the $N$th flip of the coin, where

\begin{equation}
\varphi_{n/N}\left(  p\right)  \propto p^{n/N}\left(  1-p\right)  ^{1-n/N}%
e^{-\lambda H\left(  p\right)  }\label{distribution}%
\end{equation}
and $n$ is the number of times `head' is observed among a total of $N$ flips. Because $\hat P_{N}( p)$
is obtained by raising $\varphi_{n/N}\left(  p\right)  $ to its $N$th power,
after a large number of coin flips the inference is dominated by the maxima of
$\varphi_{n/N}\left(  p\right)  $. This function can have one or two local
maxima, depending upon the value of the \textit{empirical bias}, $n/N$, and the
weigth of the cost, $\lambda$. In the limit of large $N$, $\hat P_{N}(  p)  $ converges to a
delta function centered at the inferred probability, $\hat p 
={\operatorname{argmax}}[  \varphi_{n/N}(  p)  ]  $,
which is subjected to the fluctuations in the
empirical bias, $n/N$. This inference calculation leads to qualitatively
different scenarios for a biased coin and a fair coin (we discuss the latter case in the next section).

In the case of a \textit{biased coin}, the inference always converges to a
probability, $\hat p$, controlled by (but in general not equal to) the expectation of $n/N$ (the true bias of the
coin). The value of $\hat p$ is not equal to the true bias of the coin because it also
depends upon the strength of the unpredictability cost, $\lambda$:\ the larger
$\lambda$, the more exacerbated the inferred bias. If `head' is more probable than
`tail', then the inference will enhance this bias ($\hat p>p>1/2$), and similarly if `tail' is
more probable than `head' ($\hat p <p<1/2$). For sufficiently strong regularization ($\lambda
>1$), the inferred bias is always appreciable: even for an infinitesimally
small (empirical) bias, the inferred bias remains non-vanishing. Thus, the unpredictability cost
boosts even the slightest bias present in the observations
into an appreciable bias in the inferred probabilistic origin of these observations
(Fig. \ref{fig-biased}D).

\begin{figure}[htbp]
\centering
\includegraphics[max width=0.7\textwidth]{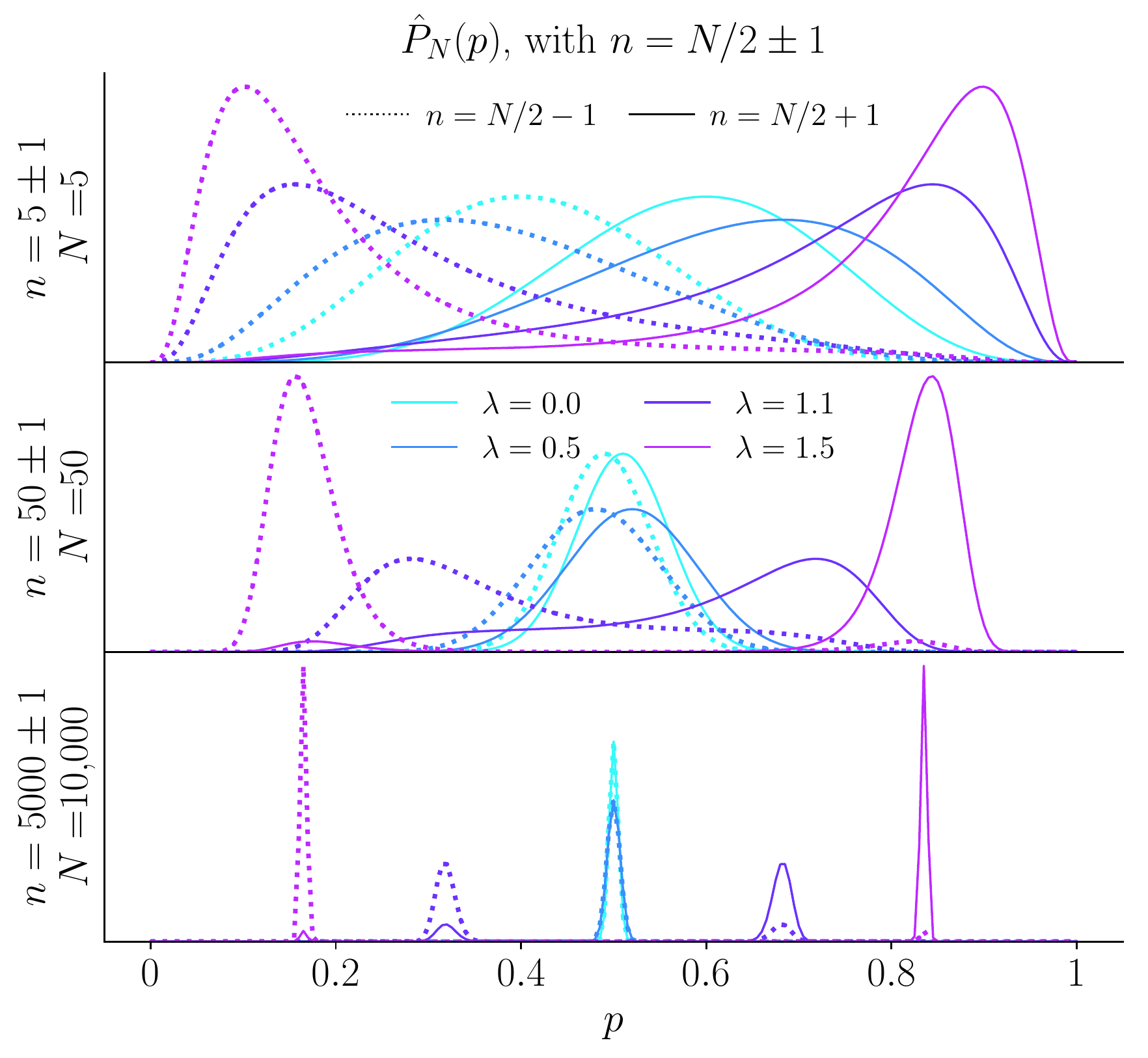}
\caption{\textbf{Inferred bias from a fair coin under the unpredictability cost.} Inferred distribution $\hat P_N(p)$ for $N=10$ (top), $100$ (middle), and $10,000$ (bottom) coin flips, when the number $n$ of `heads' is $N/2$ plus one (solid lines) or minus one (dashed line), with the unpredictability cost. Depending on the sign of $n-N/2$, the maximum of $\hat P_N(p)$ is greater or lower than 1/2. With more coin flips, for $\lambda < 1$ the two maxima (above and below 1/2) converge to 1/2. For $\lambda > 1$, the two maxima stay separated and thus the inferred probability can switch from one maximum to the other, depending on the fluctuations of $n$ around $N/2$.}
\label{fig-fair}
\end{figure}

The extreme sensitivity of the inferred probability to the empirical bias, noted above in
the context of the long-time solution (Fig. \ref{fig-biased}, C-D),
also has implications for transients in the inference. Consider a small true
bias of the coin: the probability of `head' is $1/2+\varepsilon$, with $\varepsilon\ll1$.
Typically, in half of the experiments, a fluctuation in the empirical bias
overcompensates the true bias for as long as $N\approx\varepsilon^{-2}$.
During this transient, the model subject will believe that the more likely
bias is opposed to the true one, i.e., favors `tail'. Furthermore, a model
subject with large unpredictability cost will assign a non-vanishing inferred
bias for `tail', because of its extreme sensitivity to the empirical bias.
This sensitivity yields counter-intuitive behavioral patterns when the true bias of the coin vanishes, i.e., with a fair coin: we now turn to this singular case.

\begin{figure}[htbp]
\includegraphics[max height=0.9\textheight, max width=\textwidth]{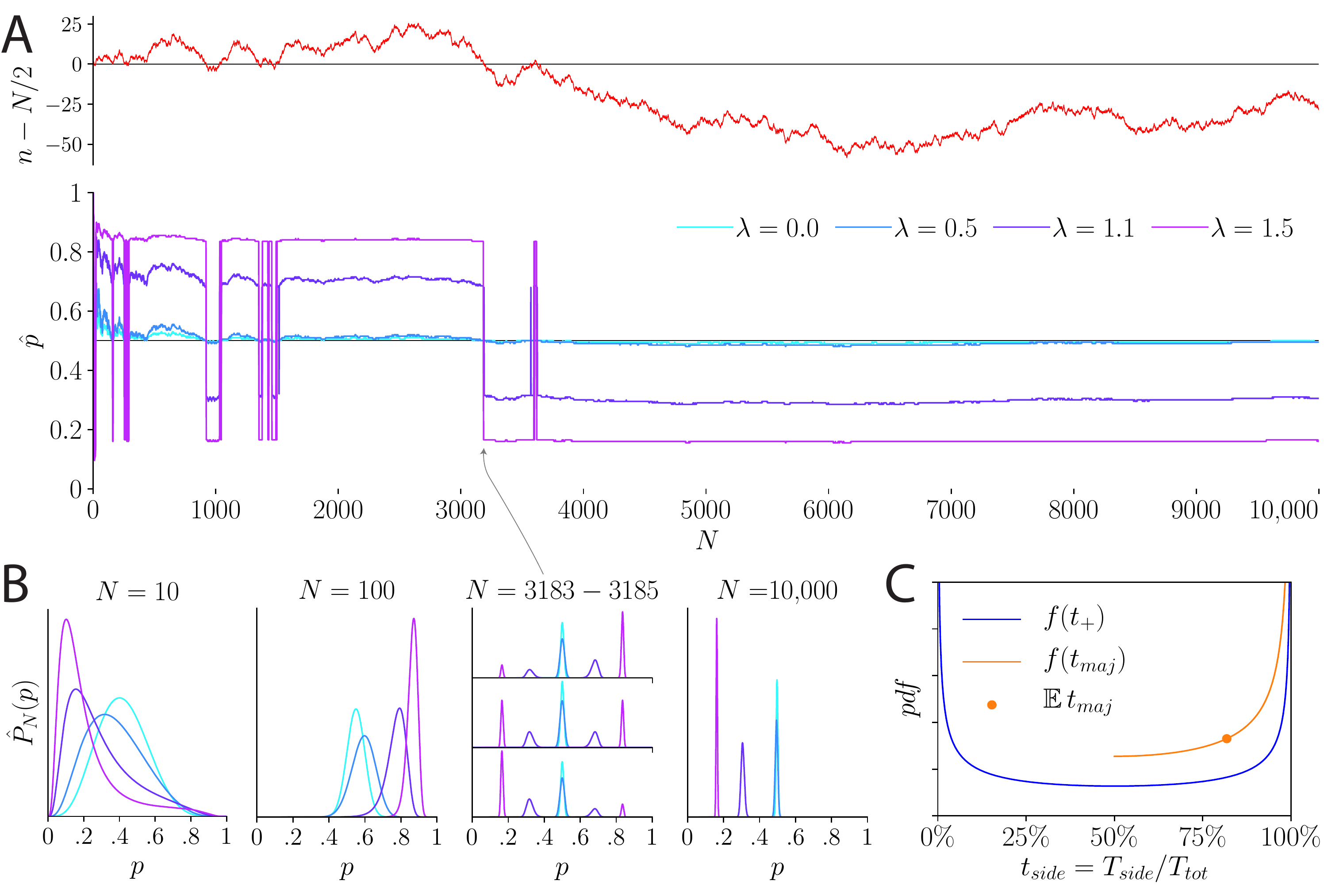}
\caption{\textbf{Behavior of the unpredictability-cost model over 10,000 flips of a fair coin.}
	\textbf{A.} \textit{Top:} Random walk described by the fluctuations of the number $n$ of `heads' around $N/2$. The quantity $n-N/2$ has excursions of varying lengths above and below zero.
			\textit{Bottom:} Inferred probability, $\hat p$, over the course of the 10,000 coin flips. Depending on the sign of $n-N/2$, the inferred probability is above or below 1/2. For $\lambda < 1$, $\hat p$ converges to 1/2, while for $\lambda > 1$ it never converges but switches between two maxima, $p_>$ and $p_<$, above and below 1/2.
	\textbf{B.} Inferred distribution, $\hat P_N(p)$, at various times during the course of the coin flips. From $N=10$ to $N=100$, the density becomes narrower; it is subject to the fluctuations of the coin flips, but for $\lambda < 1$ the density approaches 1/2. For $N=3183$ to 3185, the number of observed `heads' goes from just above $N/2$ to just below. As a result, for $\lambda > 1$ the maximum at $p_>$, above 1/2, decreases and the other maximum at $p_<$ increases, becoming the global maximum. For exactly $n=N/2$, the density is symmetric and both maxima have the same height. For $\lambda < 1$, there is only one maximum. Finally, for $N=10000$, in this instance the random walk is relatively far from 0, thus for all $\lambda>1$ the distribution is concentrated around a maximum below 1/2.
	\textbf{C.} Probability density function of the proportion of time spent on a given side --- above or below 0 --- of the random walk, and on the most visited side.
	The density of the time $T_{side}$ spent on a given side, relative to the total duration $T_{tot}$ of the random walk, follows an arcsine distribution (blue line). The extremes are more likely than the center, i.e., a random walk is not likely to spend around half of its time on a given side: it's much more likely to spend either a very long time or a very short time on this side.
	One side, the `majority' side, will thus dominate. The proportion $t_{maj}$ of time spent on this majority side has an arcsine density `folded' on values above 50\% (orange line). Its expected mean is 81.8\% (orange dot), i.e., on average unbiased random walks spend 81.8\% of their time on one side.
}
\label{fig-randomwalk}
\end{figure}

\subsection{The case of a fair coin: fluctuating and biased beliefs}

We examine the behavior of a model subject whose inference is regularized by the unpredictability cost or by the precision cost, when the coin is fair, i.e., when $p=1/2$. 
With the unpredictability cost, the case of a fair coin presents a non-trivial inference scenario
when the cost is strong ($\lambda>1$). (If
$\lambda<1$, inference converges to the true solution, with $\hat p=1/2$.)
The function $\varphi_{n/N}\left(  p\right)$ has two maxima, but now their
relative amplitudes are controlled by the \textit{fluctuations} in the empirical
bias ($n/N$), rather than by its mean (Fig. \ref{fig-fair}). It is straightforward to show that the
locations of the two maxima converge to two values, $p_{<}\left(
\lambda\right)  <1/2$ and $p_{>}\left(  \lambda\right)  >1/2$, respectively,
and that the difference of their heights follows an unbiased random walk
proportional to $n-N/2$. Whenever this random walk takes
negative values, the inferred, \textit{biased} probability of `head' is
$p_{<}\left(  \lambda\right)  $; conversely, when the random walk takes
positive values, the inferred probability of `head' is $p_{>}\left(
\lambda\right)  $ (Fig. \ref{fig-randomwalk}A,B).

In a long experiment the inferred bias never converges (if $\lambda>1$);\ it switches between
two values, and the switching times follow the non-trivial --- and rather
counter-intuitive --- statistics of return times of a random walk \cite{Feller1967}. The
distribution of the duration between successive switches in the inferred bias
is heavy-tailed, with a diverging mean. As a result, the belief that the fair
coin has a given bias lasts typically for long durations on the
order of the duration of the entire experiment. In particular, whenever the
experiment is stopped, it is likely that the model subject has maintained a
belief in an inferred bias, either $p_{<}\left(  \lambda\right)  $ or
$p_{>}\left(  \lambda\right)  $, throughout most of the experiment. On
average, a model subject believes in one of the two biased probabilities for
about $82\%$ of the duration of the experiment, no matter how long the
experiment (Fig. \ref{fig-randomwalk}C).

We emphasize that the regularization of the inference process by the
unpredictability cost does not enforce this fluctuating behavior explicitly. The
latter emerges naturally from the dynamics of the inference:~the inferred bias
is governed by the fluctuations in the stimulus, and switches between two
values without ever converging. However, because of the peculiar properties of
return times of random walks, it is much more likely to observe a small number
of such switches, in any given experiment, rather than a large number.
Typically, any given experiment is dominated by one of two inferred biases,
either $p_{<}\left(  \lambda\right)  $ or $p_{>}\left(  \lambda\right)$ (Fig. \ref{fig-randomwalk}).

\begin{figure}[htbp!]
\includegraphics[max height=0.9\textheight, max width=\textwidth]{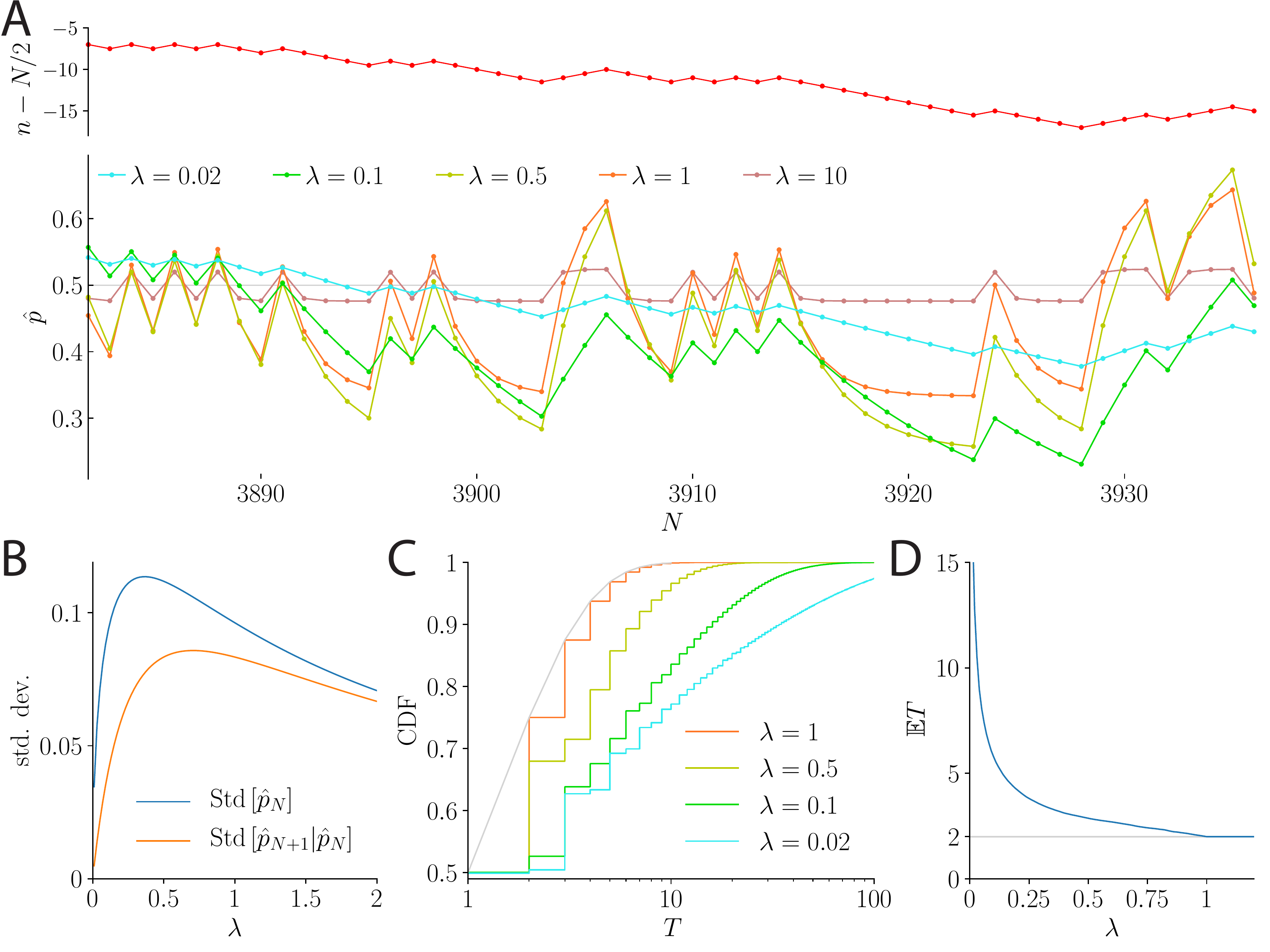}
\caption{\textbf{Behavior of the precision-cost model over 55 flips of a fair coin.}
	\textbf{A.} \textit{Top:} Random walk described by the fluctuations of the number $n$ of `heads', relative to $N/2$, over 55 successive coin flips.
		\textit{Bottom:} Inferred probability, $\hat p$, over the course of the sequence, for various values of $\lambda$. Under a weak cost ($\lambda$ small), short-term variations of the inferred probability are small, and flights away from 1/2 are long. Under stronger costs, the inferred probability is susceptible to larger changes, as it varies with the recent history of coin flips, and flights are shorter; but the overall magnitude of the fluctuations becomes smaller for larger $\lambda$.
	\textbf{B.} Standard deviation of the inferred probability, $\hat p_N$, unconditional (blue), and conditional on the preceding inferred probability (orange). The latter is also the standard deviation of the change in the inferred probability following a coin flip, conditional on the preceding inferred probability, $\operatorname{Std}[\hat p_{N+1} - \hat p_N | \hat p_N]$.
	\textbf{C.} Cumulative distribution function of the flight duration, $T$ (i.e., the number of flips during which the model subject maintains a belief in a given sign of the coin's bias), for several values of $\lambda$. 
	\textbf{D.} Expected flight duration, $\E [T]$, as a function of the strength of the cost, $\lambda$.}
\label{fig-randomwalk-leaky}
\end{figure}

The behavior of a model subject under the precision cost, in the presence of a fair coin, is qualitatively different from that under the unpredictability cost. For large $N$, the posterior fluctuates with the recent history of coin flips; its expectation, $\hat p$, follows an autoregressive process of order 1, whose two parameters are its mean, which is the correct probability, $1/2$, and its coefficient, equal to $1/(1+\lambda)$ (see Methods). The variations of the mean result from `local' variations in the empirical bias that reflect the recent history of coin flips over a timescale determined by the strength of the cost, $\lambda$.
When $\lambda$ is small, this timescale is long, and short-term variations are small in comparison to the overall variability of $\hat p$. Successive estimates are highly correlated, resulting in long `flights' away from the mean, $1/2$, and infrequent changes of the sign of the inferred bias. With larger costs, the estimate depends less on past estimates and more on recent outcomes of the coin flips, resulting in frequent changes in the belief about the sign of the coin's bias (Fig. \ref{fig-randomwalk-leaky}A,B).

More precisely, as the model subject gradually `forgets' past flips, its counts of `heads' and `tails' are exponentially filtered (Eq. (\ref{eq:n0n1})), and thus they are bounded: even with an infinite sequence of `heads', the effective count of `heads' is finite (it is the sum of a geometric series with ratio smaller than unity). As a result, the estimate of the probability, $\hat p$, is confined between two values placed symmetrically above and below $1/2$; and the distance between these two values is smaller for a strong cost, $\lambda$. If the cost is large ($\lambda\geq1$), the estimate is always close enough to $1/2$ so that just one coin flip can make it switch to the opposite inferred bias. For instance, after a long streak of `heads', the model subject infers that the coin is biased towards the `heads' ($\hat p \simeq2/3$, for $\lambda=1$); but one `tail' observation suffices to reverse this inference, in favor of a bias towards the `tails' (Fig. \ref{fig-randomwalk-leaky}A). Hence, for all $\lambda \geq 1$, the model subject is always one coin flip away from changing its belief, and this happens with probability $1/2$. Therefore, the probability that the inferred bias stays in either direction for a flight duration of exactly $T$ successive coin flips is $(1/2)^T$. As a result, the model subject holds a belief in a given sign of the bias for an average flight duration of two flips (Fig. \ref{fig-randomwalk-leaky}C, D).
For weaker costs ($\lambda<1$), longer flights are more likely. The tail of the distribution of flight durations is heavier for small values of $\lambda$, and long flights become more probable (Fig. \ref{fig-randomwalk-leaky}C; note the logarithmic scale). This results, as $\lambda$ nears zero, in large expectations of flight durations, i.e., the model subject maintains its belief in a given sign of the coin's bias for relatively long average durations (Fig. \ref{fig-randomwalk-leaky}D; the expected flight duration remains finite as long as $\lambda$ is non-vanishing \cite{Novikov2007} --- if $\lambda$ vanishes, i.e., in the optimal, Bayesian case, the inferred probability does not follow an autoregressive process, and the expected flight duration diverges).

\section{Discussion}

\subsection{Summary}
We have proposed and investigated a theoretical account of cognitive biases that emerge when humans make inferences about probabilities. The main postulate of this theory is that the brain aims at carrying out sound, Bayesian inference, but a cognitive cost --- a functional of the inferred posterior --- hinders its ability to do so. Instead of choosing the Bayesian posterior, the brain chooses another, less costly, distribution; we formalize this trade-off as ensuing from the minimization of a loss function (Eq. (\ref{eq:loss})). We consider two different costs, and examine the resulting behaviors in the case of coin flips, i.e., when a Bernoulli probability is inferred.

The \textit{precision cost} penalizes less entropic distributions (e.g., distributions concentrated around some value). Under this cost, the posterior does not converge, but fluctuates with the recent history of coin flips, and the average inferred probability is less extreme than the true probability (i.e., it is closer to 1/2). In the case of a fair coin ($p=1/2$), the model subject believes that the coin has a bias of a given sign for durations that decrease with the strength of the cost. With a strong cost ($\lambda > 1$, i.e., the weight of the cost in the loss function is greater than the weight of the distance measure), just one coin flip is enough to switch the sign of the inferred bias.
The \textit{unpredictability cost} penalizes posteriors that imply unpredictable environments. Under this cost, if the coin is biased ($p\neq1/2$), the inferred posterior converges to a probability that is more extreme than the true probability (i.e., it is further from $1/2$, and closer to 0 or to 1), and if the cost is strong, an infinitesimal true bias of the coin results in an appreciable inferred bias. In the case of a fair coin ($p=1/2$), and with a strong cost ($\lambda > 1$), the posterior does not converge, and the inferred probability abruptly switches between two values, one above and one below $1/2$, as the true empirical bias fluctuates between small positive and negative values; but the belief in a given sign of the bias typically lasts for a long duration, comparable to that of the experiment.

\subsection{Costly inference vs. erroneous beliefs}
The \textit{precision cost} yields an inference process in which past observations are gradually forgotten (Eq. (\ref{eq:n0n1})). This results from the structure of the loss function (Eq. (\ref{eq:loss})) in which the objective of learning the true value of the probability competes with the cost of being precise in doing so. The cost incites the model subject to discard, after each coin flip, some amount of information. (Another model subject bearing the same precision cost but endowed with a different objective may choose differently the information to be discarded.) This forgetfulness results from a cost that is large for narrow probability densities and, hence, can be interpreted as penalizing a form of overfitting, in line with the maximum-entropy principle  \cite{Jaynes1957,Banavar2010}.

A similar exponential forgetfulness is an essential assumption of some of the models proposed in the literature on biases in human inference; there, forgetfulness is interpreted as a rational process in the face of environments governed by statistics that vary over time.
Some experimental tasks are indeed set in such a context, as when human subjects are asked to make inferences in volatile environments \cite{Rushworth2008,Mathys2011,Gallistel2014,Khaw2017,Prat-Carrabin2021} (see also models of temporal discounting \cite{Commons1982,Commons1991,Sozou1998,Green2004,Gabaix2017}).
In such situations, older events provide less useful information than more recent ones, and, as such, forgetfulness is a way to avoid using `outdated' information. But a number of theoretical studies go beyond this view and assume that (exponential) forgetfulness occurs even in situations in which the environmental statistics are static \cite{Yu2008,Meyniel2016}. They argue that humans are so strongly adapted to changing environments that they still hold this belief even after having been exposed to a static environment over a long period of time. What such an approach is tacit about, however, is how the specific temporal structure characterizing the belief is chosen or how it may change as a function of experience. This view is to be contrasted with the framework presented here, in which there is no explicit assumption of forgetfulness; instead, a decaying memory arises as a consequence of a cognitive cost on the precision of internal representations. While we have applied our framework to the arguably simplest case of a binary, i.i.d. random variable, it may be applied to more general problems.

It is worth noting that the topic of the correctness or incorrectness of beliefs in the context of inference extends well beyond the question of whether these should include a temporal dependence. Many observed biases may be explained as resulting from a particular choice of incorrect belief or prior. Indeed, the complete class theorem \cite{Brown2007,Wald1947} shows that any set of choices is Bayes optimal under some form of prior belief. One can therefore derive from behavioral data the prior belief that makes the behavior optimal (at least in principle). This observation underlines the relevance of carrying out behavioral experiments that can cover a range of conditions and parameters, with a view to examining how putative beliefs vary and what explanation of the behavior is most parsimonious.

In the spirit of the complete class theorem, the \textit{unpredictability cost} can be interpreted as embodying the belief that predictable random processes are more likely than unpredictable ones, in the context of hierarchical inference over sources of uncertainty. A subject behaving according to our model under unpredictability cost can then, alternatively, be viewed as behaving rationally provided she holds incorrect beliefs. But if cognitive constraints \textit{do} exist, then one may not need to have recourse to assuming incorrect beliefs in order to explain biases. Furthermore, procedurally the two approaches are not equivalent: a costly inference model can predict trends in behavior as a function of variations of the experimental conditions, whereas models invoking erroneous beliefs must, in addition, specify how these may change as a function of experimental conditions.

\subsection{Costly inference vs. variational inference}
Our approach, in this study, is thus to postulate that humans approximate a Bayesian observer because of cognitive limitations.
This idea is the object of an active debate in cognitive neuroscience \cite{Penny2012,Pouget2013,Sanborn2015,Gershman2016}; it is set within a larger picture according to which structural constraints curb brain functions in perception, inference, learning, and decision-making \cite{Griffiths2015,Lieder2019,Griffiths2020,Bhui2021,Summerfield2021}.
In cognitive science, this picture can be traced back to the concept of `bounded rationality', introduced by Simon \cite{Simon1955,Simon1997}, but the proposal of a `resource-rational' approach \cite{Griffiths2015,Lieder2019,Griffiths2020} is perhaps closest to our approach, as it explicitly formalizes the mechanisms used by the brain as resulting from the optimization of a loss function equal to a negative objective plus a cost (in our case, Eq. (\ref{eq:loss}), see also Ref. \cite{Icard2015}).

What objective and what cost best characterize the loss function optimized by the brain in a given situation remains a matter of debate \cite{Ma2020}. Here, we have focused on an objective function chosen as the Kullback-Leibler divergence between the inferred distribution and the Bayesian posterior, $D_{KL}(\hat P || P)$.
This choice suggests an interesting connection with so-called `variational approaches'. In our case, the objective function is traded off with a cost. In the case of variational approaches, the often intractable optimization of the objective function is replaced by a tractable problem in which the optimization is carried out over a restricted class of (parameterized) functions \cite{Ghahramani1997,Dauwels2007,Penny2012,Gershman2016,Beal2003,Bishop2006}.
Our formulation (Eq. (\ref{eq:loss})) can thus be viewed as a variant of a variational approach: in most instances of the latter, functions belonging to a restricted class bear no cost, while all other functions come at an infinite cost; in our case, all posteriors are allowed, but incur an information-theoretic cost.
Although it is possible that the brain can only carry out inferential computations using a restricted family of distributions, we have chosen to study two `smoother' costs, that do not preclude \textit{a priori} the use of any kind of distributions. These two costs are grounded in information theory: the precision cost varies with the entropy of the posterior, while the unpredictability cost depends on the entropy of the inferred probabilities.

The connection of our framework with variational inference is particularly relevant in the context of studies that rely on the minimization of a quantity that has become known as ‘evidence bound’ in machine learning \cite{Winn2005} and `free energy' in cognitive science \cite{Friston2010}.
This quantity is equal to the sum of the Kullback-Leibler divergence between the inferred posterior and the Bayesian posterior, $D_{KL}(\hat P || P)$, i.e., the objective function in Eq. (\ref{eq:loss}), and a term that does not depend upon the inferred posterior. Thus, the minimization of the free energy, in these studies, amounts to the minimization of our objective function; but it is carried out within a restricted class of posteriors in the case of variational inference, while in our case the objective function is regularized by a cost.
In an alternative formulation, the free energy can be written as the sum of two terms, the entropy of the inferred posterior, i.e., our precision cost, and an 'energy' term.
Optimization then becomes a version of the maximum-entropy principle \cite{Jaynes1957} in which the energy term (in our case, the first term on the right-hand-side of Eq. (\ref{eq:loss})) acts as a constraint or regularizer, while the entropy is maximized. This interpretation of Eq. \mbox{(\ref{eq:loss})} is dual to the one we have entertained throughout the paper, namely that entropy is used as a regularizer.

\subsection{Costly inference and human behavior}
Our goal, in this study, was not to establish the extent to which each model captures the behavior of human subjects quantitatively. We note, however, that recent models of human inference are based on exponentially filtered counts of the outcomes \cite{Yu2008,Meyniel2016}, precisely of the kind that result from our precision cost (Eq. (\ref{eq:n0n1})). In addition, this cost predicts that inferred probabilities are less extreme than the true probabilities \mbox{(Fig.~\ref{fig-biased}B)}, a behavior that was reported in experiments of decision-making under risk with human subjects \cite{Gonzalez1999,Zhang2020}. Conversely, in decisions from experience, subjects were found to underestimate the probability of rare events \cite{Hertwig2004}; the unpredictability cost predicts a bias in this direction (Fig. \ref{fig-biased}D).

The different predictions of our models call for a detailed and quantitative examination of the behavior of human subjects in inference tasks. Most existing empirical studies of human inference of probabilities focus on the singular case of an equiprobable environment ($p=1/2$) \cite{Cho2002,Yu2008,Gokaydin2017}. The models, however, make predictions about how the biases in inference should vary with the Bernoulli probability of outcomes, $p$. This begs for the design of a more flexible experimental paradigm, in which human behavior is investigated in environments that reflect a range of Bernoulli probabilities or, more broadly, an array of different generative statistics. This kind of experimental investigation, which addresses behavioral biases quantitatively and over a broad range of conditions, is warranted not only in the context of fundamental research but may prove useful also in the nascent field of computational psychiatry. It would provide a basis to study and compare the structure of cognitive processes in patients and healthy subjects \cite{Stephan2014,Adams2016,Schwartenbeck2016,Ashinoff2021}.

\section{Methods}\label{sec:Methods}

\subsection{Precision cost: variance of the posterior}
We derive an approximation of the variance of the posterior inferred by the model subject under a precision cost with strength $\lambda$. The posterior at trial $N$ is a Beta distribution with parameters $\hat n_1$ and $\hat n_0$ (Eqs. (\ref{eq:beta_posterior}, \ref{eq:n0n1})). Its variance is
\begin{equation}
\operatorname{Var}[\hat P_N] = \frac{(\hat n_1 + 1)(\hat n_0 + 1)}{(\hat n_1 + \hat n_0 + 2)^2 (\hat n_1 + \hat n_0 + 3)}.
\end{equation}
For $N$ large, we can calculate the expected value of this variance, as
\begin{equation}
\E \operatorname{Var}[\hat P_N] = \lambda
	\frac	{ 4p(1-p) + \lambda(1+\lambda)(2+\lambda) }
	  	{ (1+2\lambda)^2 (1+3\lambda) (2+\lambda)},
\end{equation}
where $p$ is the true value of the probability.
For small $\lambda$, we obtain the approximation
\begin{equation}
\E \operatorname{Var}[\hat P_N] \simeq \lambda 4 p (1-p),
\end{equation}
while for large $\lambda$, we obtain
\begin{equation}
\E \operatorname{Var}[\hat P_N] \simeq \frac{1}{12} \Big(1-\frac{1}{3\lambda} \Big).
\end{equation} 
Note that $1/12$ is the variance of the uniform distribution on the interval $[0,1]$.

\subsection{Precision cost: autoregressive process}
We show, here, that for large $N$ the inference under the precision cost results in the inferred probability following an autoregressive process. At trial $N$, the mean, $\hat p_N$, of the posterior, $\hat P_N(p)$, defined in Eq. (\ref{eq:beta_posterior}), is a function of the exponentially filtered counts, $\hat n_0$ and $\hat n_1$ (Eq. (\ref{eq:n0n1})), as
\begin{align}
\hat p_N &= \frac{\hat n_1 + 1}{\hat n_0 + \hat n_1 + 2} \\
   &= \frac{\sum_{k=0}^{N-1} (\frac{1}{1+\lambda})^{k+1} x_{N-k} + 1}{\sum_{k=0}^{N-1} (\frac{1}{1+\lambda})^{k+1} + 2}.
\end{align}
Taking the expectation of this expression, we obtain, for large $N$, the expected inferred probability, as

\begin{equation}
\E [ \hat p ] = \frac{1}{1+2\lambda} p + \frac{2\lambda}{1 + 2\lambda} \frac{1}{2},
\end{equation}
an intermediate value between 1/2 and the true value of the probability, $p$. Thus, in this model, the inferred probabilities are less `extreme' than the true probabilities (Fig. \ref{fig-biased}B). Furthermore, we can express the inferred probability at trial $N+1$ as a function of the inferred probability at trial $N$. We obtain, for large $N$,

\begin{equation}
\hat p_{N+1} = \frac{\lambda}{1+\lambda} \E [ \hat p ] + \frac{1}{1+\lambda} \hat p_N
			+ \frac{\lambda}{(1+\lambda)(1+2\lambda)} (x_{N+1} - p),
\end{equation}
which defines an autoregressive process of order 1 with coefficient $1/(1+\lambda)$.

\bibliographystyle{unsrt}
\bibliography{References-Inference}

\section*{Funding}
This work was supported by the CNRS through UMR8023, the Global Scholar Program at Princeton University, and the Visiting Faculty Program at the Weizmann Institute of Science. A.P.C. was supported by a PhD fellowship of the Fondation Pierre-Gilles de Gennes pour la Recherche.

\section*{Conflict of interest}
The authors declare no conflict of interest.

\end{document}